\documentclass[11pt]{article}

\begin{document}

\title{\bf World Spinors: \\
Group Representations and Field Equations}

\author{{\bf Djordje \v Sija\v cki} \\ Institute of Physics, P.O. Box 57,
11001 Belgrade, Serbia}

\date{}

\maketitle

\begin{abstract}
The questions of the existence, basic algebraic properties and
relevant constraints that yield a viable 
physical interpretation of world spinors are discussed in
details. Relations between spinorial wave equations that
transform respectively w.r.t. the tangent flat-space (anholonomic)
Affine symmetry group and the world generic-curved-space (holonomic)
group of Diffeomorphisms are presented. A geometric construction based
on an infinite-component generalization of the frame fields
(e.g. tetrads) is outlined. The world spinor field equation in $3D$ is
treated in more details.
\end{abstract}

\vskip12pt

\section{Introduction.}

The Dirac equation turned out to be one of the most successful equations
of the XX century physics - it describes the basic matter constituents
(both particles and fields), and very significantly, it paved a way to
develop the concept of gauge theories thus completing the description of
the basic interactions as well. It is a Poincar\'e invariant linear field
equation which describes relativistic spin $\frac{1}{2}$ particles, with
interactions naturally introduced by the minimal coupling prescription.

Here we go {\bf beyond the Poincar\'e invariance} and study
{\bf affine invariant generalizations} of the Dirac equation. In other
words, we consider a {\bf generalization that will describe a
spinorial  field - world spinors - in a generic curved spacetime}
$(L_{n}, g)$, {\bf characterized by arbitrary torsion and
general-linear curvature}. Note that the spinorial fields in the
non-affine generalizations of GR (which are based on
higher-dimensional orthogonal-type generalizations of the Lorentz
group) are only allowed for special spacetime configurations and fail
to extend to the generic case.

\vskip24pt The {\bf finite-dimensional world tensor fields} in $R^{n}$
are characterized by the non-unitary irreducible representations of
the general linear subgroup $GL(n,R)$ of the Diffeomorphism group
$Diff(n,R)$. In the flat-space limit they split up into $SO(1,n-1)$
($SL(2,C)/Z_2$ for $n=4$) irreducible pieces. The corresponding
particle states are defined in the tangent flat-space only. They are
characterized by the unitary irreducible representations of the
(inhomogeneous) Poincar\'e group $P(n) = T_n\wedge SO(1,n-1)$, and
they are defined by the relevant "little" group unitary representation
labels.

In the {\bf generalization to world spinors}, the double covering group,
$\overline{SO}(1,n-1)$, of the $SO(1,n-1)$ one, that characterizes a
Dirac-type fields in $D=n$ dimensions, is enlarged to the
$\overline{SL}(n,R) \subset \overline{GL}(n,R)$ group,
$$
\overline{SO}(1,n-1) \quad\mapsto\quad \overline{SL}(n,R) \subset
\overline{GL}(n,R)
$$
while $SA(n,R) =
T_n\wedge \overline{SL}(n,R)$ is to replace the Poincar\'e group itself.
$$
P(n) \quad\mapsto\quad \overline{SA}(n,R) = T_n\wedge \overline{SL}(n,R)
$$

{\bf Affine "particles"} are characterized by the unitary irreducible
representations of the $\overline{SA}(n,R)$ group, that are actually
nonlinear unitary representations over an appropriate "little" group.
E.g. for $m\ne 0$:
$$
T_{n-1}\otimes \overline{SL}(n-1,R) \supset T_{n-1}\otimes
\overline{SO}(n)
$$

A mutual {\bf particle--field correspondence} is achieved by requiring
{\bf (i)} that fields have appropriate mass (Klein-Gordon-like equation
condition, for $m\ne 0$), and {\bf (ii)} that  the subgroup of the
field-defining homogeneous group, which is isomorphic to the homogeneous
part of the "little" group, is represented unitarily. Furthermore, one
has to project away all representations except the one that characterizes
the particle states.

A {\bf physically correct picture, in the affine case, is
obtained by making use of the $\overline{SA}(n,R)$ group unitary
(irreducible) representations for "affine" particles.}
{\it The affine-particle states are characterized by the unitary
(irreducible) representations of the $T_{n-1}\otimes
\overline{SL}(n-1,R)$ "little" group. The intrinsic part of these
representations is} {\bf necessarily infinite-dimensional} {\bf due to
non-compactness of the $SL(n,R)$ group.}
{\it The corresponding affine fields should be described by the
non-unitary infinite-dimensional $\overline{SL}(n,R)$ representations,
that are} {\bf unitary} {\it when restricted to the homogeneous "little"
subgroup} $\overline{SL}(n-1,R)$.
Therefore, {\bf the first step towards world spinor fields is a
construction of infinite-dimensional non-unitary $\overline{SL}(n,R)$
representations, that are unitary when restricted to the
$\overline{SL}(n-1,R)$ group. These fields reduce to an infinite sum of
(non-unitary) finite-dimensional $\overline{SO}(1,n-1)$ fields}.

\section{Existence of the double-covering $\overline{GL}(n,R)$.}

Let us state first some relevant mathematical results.\\
{\it Theorem 1:}
Let $g_{0}=k_{0}+a_{0}+n_{0}$ be an Iwasawa decomposition of a semisimple
Lie algebra $g_{0}$ over $R$. Let $G$ be any connected Lie group with Lie
algebra $g_{0}$, and let $K,A,N$ be the analytic subgroups of $G$ with
Lie algebras $k_{0}$,$a_{0}$ and $n_{0}$ respectively. The mapping
$(k,a,n)\rightarrow kan$ $(k\in K,a\in A,n\in N)$
is an analytic diffeomorphism of the product manifold $K\times A\times N$
onto $G$, and {\it the groups $A$ and $N$ are simply connected.}

Any semisimple Lie group can be decomposed into the product of the {\it
maximal compact subgroup} $K$, an {\it Abelian group} $A$ and a {\it
nilpotent group} $N$. {\bf As a result of Theorem 1, only $K$ is not
guaranteed to be simply-connected}. There exists a universal covering
group $\overline{K}$ of $K$, and thus also a universal covering of $G$:
$\overline{G} \simeq \overline{K} \times  A \times  N.$

For the group of diffeomorphisms, let $Diff(n,R)$ be the group of all
homeomorphisms $f$ or $R^{n}$ such that $f$ and $f^{-1}$ are of class
$C^{1}$. Stewart proved the decomposition
$Diff(n,R) = GL(n,R) \times H \times R_{n}$,
where the subgroup $H$ is contractible to a point. Thus, as $O(n)$
is the compact subgroup of $GL(n,R)$, one finds

{\it Theorem 2:}
$O(n)$ is a deformation retract of $Diff(n,R)$.

As a result, there exists a universal covering of the Diffeomorphism
group
$\overline{Diff}(n,R) \simeq \overline{GL}(n,R) \times H \times
R_{n}.$

{\bf Summing up, we note that both $SL(n,R)$ and on the other hand
$GL(n,R)$ and $Diff(n,R)$ will all have double coverings, defined by
$\overline{SO}(n)$ and $\overline{O}(n)$ respectively, the
double-coverings of the $SO(n)$ and $O(n)$ maximal compact subgroups}.

In the physically most interesting case $n=4$, there is a
homomorphism between $SO(3)\times SO(3)$ and $SO(4)$. Since $SO(3)
\simeq SU(2)/Z_{2}$, where $Z_{2}$ is the two-element center $\{
1,-1\}$, one has $SO(4) \simeq [SU(2)\times SU(2)]/Z_{2}^{d}$, where
$Z_{2}^{d}$ is the diagonal discrete group whose representations are
given by $\{ 1,(-1)^{2j_{1}}=(-1)^{2j_{2}} \}$ with $j_{1}$ and
$j_{2}$ being the Casimir labels of the two $SU(2)$ representations. The
full $Z_{2}\times Z_{2}$ group, given by the representations $\{
1,(-1)^{2j_{1}} \} \otimes \{ 1,(-1)^{2j_{2}} \}$, is the center of
$\overline{SO}(4)=SU(2)\times SU(2)$, which is thus the
quadruple-covering of $SO(3)\times SO(3)$ and a double-covering of
$SO(4)$. $SO(3)\times SO(3)$, $SO(4)$ and
$\overline{SO}(4)=SU(2)\times SU(2)$ are thus the maximal compact
subgroups of $SO(3,3)$, $SL(4,R)$ and $\overline{SL}(4,R)$
respectively. One can sum up these results by the following exact
sequences 
$$
\matrix{
   & &  &  &     1 &  &            1 &  &  \cr
   &  &  &  & \downarrow &  & \downarrow &  &  \cr
 1 & \rightarrow & Z_{2}^{d} & \rightarrow & Z_{2}\times Z_{2} &
\rightarrow & Z_{2} & \rightarrow & 1 \cr
   &  &  &  & \downarrow &  & \downarrow &  &  \cr
 1 & \rightarrow & Z_{2}^{d} & \rightarrow & \overline{SL}(4,R) &
\rightarrow & SL(4,R) & \rightarrow & 1 \cr
   &  &  &  & \downarrow &  & \downarrow &  &  \cr
   &  &  &  & SO(3,3) &  & SO(3,3) &  &  \cr
   &  &  &  & \downarrow &  & \downarrow &  &  \cr
   &  &  &  & 1 &  & 1 &  &  \cr}
$$

\section{The $SL(4,R)\rightarrow SL(4,C)$ embedding.} 

A glance at the classical semisimple Lie group Dynkin diagrams tells us
that we need to investigate two possibilities: either one can embed the
$\overline{SL}(4,R)$ algebra $sl(4,R)$ in the Lie algebra of the
appropriate noncompact version of the orthogonal algebra $so(6)$ of the
$Spin(6)$ group, or in the $sl(4,C)$ algebra of the $SL(4,C)$ group. In
the first case, the appropriate noncompact group is $Spin(3,3) \simeq
\overline{SO}(3,3)$ which is {\it isomorphic} to the $SL(4,R)$ group
itself. As for the second option, we consider first the problem of
embedding the algebra $sl(4,R) \rightarrow sl(4,C)$.

The maximal compact subalgebra $so(4)$ of the $sl(4,R)$ algebra is
embedded into the maximal compact subalgebra $su(4)$ of the $sl(4,C)$
algebra.
$$
\matrix{ sl(4,R)& \rightarrow & sl(4,C) \cr
           \cup &             &  \cup   \cr
         so(4)  & \rightarrow &  su(4)  \cr}
$$

There are two principally different ways to carry out the $so(4)
\rightarrow su(4)$.

\noindent{\bf Natural $({1\over 2},{1\over 2})$ embedding}

In this embedding the $so(4)$ algebra is represented by the genuine
$4\times 4$ orthogonal matrices of the 4-vector $({1\over 2},{1\over 2})$
representation (i.e. antisymmetric matrices multiplied by the imaginary
unit). The $SL(4,C)$ generators split with respect to the naturally
embedded $so(4)$ algebra as follows
$$
31 \supset 1_{nc} \oplus 6_{c} \oplus 6_{nc} \oplus 9_{c} \oplus 9_{nc},
$$
where $c$ and $nc$ denote the compact and noncompact operators
respectively. The $6_{c}$ and $9_{nc}$ parts generate the $SL(4,R)$
subgroup of the $SL(4,C)$ group. The maximal compact subgroup $SO(4)$ of
the $SL(4,R)$ group is realized in this embedding through its (single
valued) vector representation $({1\over 2},{1\over 2})$. In order to
embed $\overline{SL}(4,R)$ into $SL(4,C)$, one would now have to embed
its maximal compact subgroup $SO(4)$ in the maximal compact subgroup
$SL(4,C)$, namely $SU(4)$. However, that is impossible $({1\over 2},
{1\over 2})$ $\longrightarrow {\!\!\!\!\!\!\!\!{/}}$ $({1\over
2},0)\oplus (0,{1\over 2})$. The alternative could have been to embed
$\overline{SO}(4)$ in a hypothetical double covering of $SU(4)$ - except
that $SU(4)$ is simply connected and thus is its own universal covering.
We, therefore, conclude that in the {\it natural} embedding one can embed
$SL(4,R)$ in $SL(4,C)$ {\it but not the} $\overline{SL}(4,R)$ {\it
covering group}.

\noindent{\bf Dirac $\{ ({1\over 2},0)\oplus (0,{1\over 2})\}$ embedding}

In this case the $so(4)\rightarrow su(4)$ embedding is realized through a
direct sum of $2\times 2$ complex matrices, i.e.
$$
so(4) \rightarrow \left( \matrix{ su(2) &   0   \cr
                                    0   & su(2) \cr} \right) \in su(4).
$$
The $SL(4,C)$ generators now split with respect to the $su(2) \oplus
su(2)$ algebra as follows
$$
31 \supset 1_{c} \oplus 1_{nc} \oplus 1_{nc} \oplus 4_{c} \oplus 4_{c}
\oplus 4_{nc} \oplus 4_{nc} \oplus 6_{c} \oplus 6_{nc}.
$$
It is obvious from this decomposition that in the $sl(4,C)$ algebra {\it
there exist no 9-component noncompact irreducible tensor-operator} with
respect to the chosen $su(2) \oplus su(2) \sim so(4)$ subalgebra, which
would, together with the $6_{c}$ operators, form an $sl(4,R)$ algebra.
Thus, we conclude that this type of $so(4)$ embedding into $su(4) \subset
sl(4,C)$ does not extend to an embedding of either $SL(4,R)$ of
$\overline{SL}(4,R)$ into $SL(4,C)$:

To sum up, we have demonstrated that {\bf there exist no
finite-dimensional faithful representations of} $\overline{SL}(4,R)$.

\section{The deunitarizing automorphism.}

The unitarity properties, that ensure correct physical characteristics
of the affine fields, can be achieved by combining the unitary
(irreducible) representations and the so called "deunitarizing"
automorphism of the $\overline{SL}(n,R)$ group.

The commutation relations of the $\overline{SL}(n,R)$ generators
$Q_{ab}$, $a,b$ $=$ \hfill\break $0, 1, \dots , n-1$ are
$$
[Q_{ab},Q_{cd}] = i(\eta_{bc}Q_{ad} - \eta_{ad}Q_{cb}) ,
$$
taking $\eta_{ab} = diag(+1, -1, \dots , -1)$. The important subalgebras
are as follows.

{\bf (i)} $so(1,n-1)$: The $M_{ab} = Q_{[ab]}$  operators generate the
Lorentz-like subgroup $\overline{SO}(1,n-1)$ with $J_{ij} = M_{ij}$
(angular 
momentum) and $K_{i} = M_{0i}$ (the boosts) $i,j = 1,2,\dots ,n-1$.

{\bf (ii)} $so(n)$: The $J_{ij}$ and $N_{i} = Q_{\{0i\}}$ operators
generate the maximal compact subgroup $\overline{SO}(n)$.

{\bf (iii)} $sl(n-1)$: The $J_{ij}$ and $T_{ij} = Q_{\{ij\}}$ operators
generate the subgroup $\overline{SL}(n-1,R)$ - the "little" group of the
massive particle states.

{\bf The $\overline{SL}(n,R)$ commutation relations are invariant under
the ``deunitarizing'' automorphism},
\begin{eqnarray*}
 &J^\prime_{ij} = J_{ij}\ ,\quad K^\prime_{i} = iN_{i}\ ,\quad
N_{i}^\prime = iK_{i}\ ,\\
 &T^\prime_{ij} = T_{ij}\ ,\quad
T^\prime_{00} = T_{00} \ (= Q_{00})\ , \\
\end{eqnarray*}
so that $(J_{ij},\ iK_{i})$ generate the new compact
$\overline{SO}(n)^\prime$ and $(J_{ij},\ iN_{i})$ generate
$\overline{SO}(1,n-1)^\prime$.

For the massive (spinorial) particle states we use the basis vectors
of the unitary irreducible representations of
$\overline{SL}(n,R)^\prime$, so that the compact subgroup finite
multiplets correspond to $\overline{SO}(n)^\prime$: $(J_{ij},\
iK_{i})$ while $\overline{SO}(1,n-1)^\prime$: $(J_{ij},\ iN_{i})$ is
represented by unitary infinite-dimensional representations. We now
perform the inverse transformation and return to the unprimed
$\overline{SL}(n,R)$ for our physical identification:
$\overline{SL}(n,R)$ is represented non-unitarily, the compact
$\overline{SO}(n)$ is represented by non-unitary infinite
representations while the Lorentz group is represented by non-unitary
finite representations. These finite-dimensional non-unitary Lorentz
group representations are necessary in order to ensure a correct
particle interpretation (i.g. boosted proton remains proton).  Note
that  $\overline{SL}(n-1,R)$, the stability subgroup of
$\overline{SA}(n,R)$, is represented unitarily.

\section{World spinor field transformations.}

The world spinor fields transform w.r.t. $\overline{Diff}(n,R)$ as
follows

$(D(a,\bar f)\Psi_M) (x)
= (D_{ \overline{Diff}_0(n,R)})^N_M (\bar f) \Psi_N (f^{-1}(x-a))$,

$(a,\bar f) \in T_{n} \wedge \overline{Diff}_0(n,R)$,

\noindent where $\overline{Diff}_0(n,R)$ is the homogeneous part of
$\overline{Diff}(n,R)$, and $D_{\overline{Diff}_0(n,R)}$ 
$\supset$ ${\sum}^\oplus D_{\overline{SL}(n,R)}$ is the corresponding
representation in the space of world spinor field components.  As a
matter of fact, we consider here those representations of
$\overline{Diff}_0(n,R)$ that are nonlinearly realized over the
maximal linear subgroup $\overline{SL}(n,R)$ (here given in terms 
of infinite matrices).

The affine "particle" states transform according to the following
representation 
$$ 
D(a, \bar s) \rightarrow e^{i(sp)\cdot a}
D_{\overline{SL}(n,R)} (L^{-1}(sp)\bar s L(p)) ,\quad (a,\bar s) \in
T_n\wedge \overline{SL}(n,R),
$$
where $L \in \overline{SL}(n,R)/\overline{SL}(n-1,R)$, and $p$ is the
$n$-momentum. The unitarity properties of various representations in
these expressions is as described in the previous section.

Provided the relevant $\overline{SL}(n,R)$ representations are known, one
can first define the corresponding general/special Affine spinor fields
in the tangent to $R^{n}$, and than make use of the infinite-component
pseudo-frame fields $E^{A}_{M}(x)$, "alephzeroads", that generalize the
tetrad fields of $R^{4}$. Let us define a pseudo-frame $E^{A}_{M}(x)$
s.t.
$$
\Psi_{M}(x) = E^{A}_{M}(x) \Psi_{A}(x),
$$
where $\Psi_{M}(x)$ and $\Psi_{A}(x)$ are the world (holonomic), and
general/special Affine spinor fields respectively. The $E^{A}_{M}(x)$
(and their inverses $E^{M}_{A}(x)$) are thus infinite matrices related to
the quotient $\overline{Diff}_{0}(n,R)/\overline{SL}(n,R)$. Their
infinitesimal transformations are
$$
\delta E^{A}_{M}(x) = i\epsilon^{a}_{b}(x) \{Q_{a}^{b}\}^{A}_{B}
E^{B}_{M}(x) +
\partial_\mu \xi^{\nu} e^{a}_{\nu}e^{\mu}_{b}\{Q^{a}_{b}\}^{A}_{B}
 E^{B}_{M}(x),
$$
where  $\epsilon^{a}_{b}$ and $\xi^\mu$ are group parameters of
$\overline{SL}(n,R)$ and \hfill\break
$\overline{Diff}(n,R)/\overline{Diff}_{0}(n,R)$ respectively, while
$e^{a}_{\nu}$ are the standard $n$-bine fields.

The infinitesimal transformations of the world spinor fields themselves are
given as follows:
$$
\delta \Psi^{M}(x) = i\big\{ \epsilon^{a}_{b}(x) E^{M}_{A}(x)
\big(Q_{a}^{b}\big)^{A}_{B} E^{B}_{N}(x) + \xi^{\mu} \big[
\delta^{M}_{N}\partial_{\mu} +
E^{M}_{B}(x)\partial_{\mu}E^{B}_{N}(x)\big] \big\} \Psi^{N}(x).
$$
The $\big( Q^{b}_{a} \big)^{M}_{N} = E^{M}_{A}(x)
\big(Q_{a}^{b}\big)^{A}_{B} E^{B}_{N}(x)$ is the holonomic form of the
$\overline{SL}(n,R)$ generators given in terms of the corresponding
anholonomic ones in the space of spinor fields $\Psi_{M}(x)$ and
$\Psi_{A}(x)$ respectively.

The above outlined construction allows one to define a fully
$\overline{Diff}(n,R)$ covariant Dirac-like wave equation for the
corresponding world spinor fields provided a Dirac-like wave equation for
the $\overline{SL}(n,R)$ group is known. In other words, one can lift an
$\overline{SL}(n,R)$ covariant equation of the form
$$
\big( ie^{\mu}_{a} \big(X^{a}\big)^{B}_{A} \partial_{\mu} - M \big)
\Psi_{B}(x) = 0,
$$
to a $\overline{Diff}(n,R)$ covariant equation
$$
\big( ie^{\mu}_{a} E^{N}_{B} \big(X^{a}\big)^{B}_{A}E^{A}_{M}
\partial_{\mu} - M \big) \Psi_{N}(x) = 0,
$$
provided a spinorial $\overline{SL}(n,R)$ representation for the
$\Psi$ field is given, with the corresponding representation Hilbert space
invariant w.r.t. $X^{a}$ action. {\bf Thus, the crucial step towards a
Dirac-like world spinor equation is a construction of the corresponding
$\overline{SL}(n,R)$ wave equation}.

\section{$\overline{SL}(4,R)$ vector operator $X$.}

For the construction of a Dirac-type equation, which is to be invariant
under (special) affine transformations, we have two possible approaches to
derive the matrix elements of the generalized Dirac matrices $X_a$.

We can consider the defining commutation relations of a
$\overline{SL}(4,R)$ vector operator $X_a$,
\begin{eqnarray*}
&&[M_{ab}, X_c] = i \eta_{bc} X_a - i \eta_{ac} X_b \\ 
&&[T_{ab}, X_c] = i \eta_{bc} X_a + i \eta_{ac} X_b
\end{eqnarray*}

One can obtain the matrix elements of the generalized Dirac matrices
$X_a$ by solving these relations for $X_a$ in the Hilbert space of a
suitable representation of $\overline{SL}(4,R)$.
Alternatively, one can embed $\overline{SL}(4,R)$ into
$\overline{SL}(5,R)$. Let the generators of $\overline{SL}(5,R)$ be
$R_A{}^B$, $A,B = 0,...,4$. Now, there are two natural $\overline{SL}(4,R)$
four-vectors $X_a$, and $Y_a$ defined by
\begin{eqnarray*}
X_a:=R_{a4}, \quad Y_a:=L_{4a}, \qquad a=0,1,2,3 .
\end{eqnarray*}
The operator $X_a$ ($Y_a$) obtained in this way fulfills the required
$SL(4,R)$ four-vector commutation relations by construction. It is
interesting to point out that the operator $G_a=\frac{1}{2}(X_a-Y_a)$
satisfies
\begin{eqnarray*}
[G_a, G_b] = -i M_{ab} ,
\end{eqnarray*}
thereby generalizing a property of Dirac's $\gamma$-matrices. Since
$X_a$, $M_{ab}$ and $T_{ab}$ form a closed algebra, the action of
$X_a$ on the $\overline{SL}(4,R)$ states does not lead out of the
$\overline{SL}(5,R)$ representation Hilbert space.

In order to obtain an impression about the general structure of the
matrix $X_a$, let us consider the following embedding of three
finite-dimensional tensorial $SL(4,R)$ irreducible representations
into the corresponding one of $SL(5,R)$,
\begin{eqnarray*}
SL(5,R) &\supset& SL(4,R) \nonumber\\ \underbrace{ \stackrel{15}{
\begin{picture}(18,8)
\put(1,-1){\framebox(16,8)} \put(9,-1){\line(0,1){8}}
\end{picture}} }_{\varphi_{AB}}
&\supset& \underbrace{ \stackrel{10}{
\begin{picture}(18,8)
\put(1,-1){\framebox(16,8)} \put(9,-1){\line(0,1){8}}
\end{picture} } }_{\varphi_{ab}}
\oplus \underbrace{ \stackrel{4}{
\begin{picture}(18,8)
\put(1,-1){\framebox(16,8)} \put(9,-1){\line(0,1){8}}$\times$
\end{picture}} }_{\varphi_{a}}
\oplus \underbrace{ \stackrel{1}{
\begin{picture}(18,8)
\put(1,-1){\framebox(16,8)} \put(9,-1){\line(0,1){8}}$\times$
\put(-1,0){$\times$}
\end{picture}} }_{\varphi} \,,
\end{eqnarray*}
where "box" is the Young tableau for an irreducible vector representation
of $SL(n,R)$, $n=4,5$. The effect of the action of the $SL(4,R)$
vector $X_a$ on the fields $\varphi$, $\varphi_{a}$ and $\varphi_{ab}$
is 
$$
\stackrel{X_a}{
\begin{picture}(10,8)
\put(1,-1){\framebox(8,8)}
\end{picture}}
\otimes \stackrel{\varphi}{
\begin{picture}(18,8)
\put(1,-1){\framebox(16,8)} \put(9,-1){\line(0,1){8}}$\times$
\put(-1,0){$\times$}
\end{picture} }
\ \mapsto\  \stackrel{\varphi_a}{
\begin{picture}(10,8)
\put(1,-1){\framebox(8,8)}
\end{picture}}, \quad\quad
\stackrel{X_a}{
\begin{picture}(10,8)
\put(1,-1){\framebox(8,8)}
\end{picture}}
\otimes \stackrel{\varphi_a}{
\begin{picture}(18,8)
\put(1,-1){\framebox(16,8)} \put(9,-1){\line(0,1){8}} {$\times$}
\end{picture} }
\ \mapsto\  \stackrel{\varphi_{ab}}{
\begin{picture}(18,8)
\put(1,-1){\framebox(16,8)} \put(9,-1){\line(0,1){8}}
\end{picture}}, \quad\quad
\stackrel{X_a}{
\begin{picture}(10,8)
\put(1,-1){\framebox(8,8)}
\end{picture}}
\otimes \stackrel{\varphi_{ab}}{
\begin{picture}(18,8)
\put(1,-1){\framebox(16,8)} \put(9,-1){\line(0,1){8}}
\end{picture} }
\ \mapsto\  0 \,.
$$
Other possible Young tableaux do not appear due to the closure of the
Hilbert space. Gathering these fields in a vector $\varphi_M=(\varphi,
\varphi_a, \varphi_{ab})^{\rm T}$, we can read off the structure of
$X_a$,
\begin{eqnarray*}
X_a= \left[
\begin{tabular}{c|c|c}
  $0$  &  & \\
\hline
   \begin{picture}(8,30)
\multiput(0,0)(0,7){4}{x}
\end{picture} & \,\raisebox{1.4ex}{${\bf 0}_4$}  & \\
\hline
     &
\begin{picture}(23,72)
\multiput(0,0)(0,7){10}{xxxx}
\end{picture}
 &  \begin{picture}(72,72)
\multiput(0,0)(0,7){10}{} \put(30,30){${\bf 0}_{10}$}
\end{picture}
\end{tabular}
\right]\,.
\end{eqnarray*}
It is interesting to observe that $X_a$ has zero matrices on the
block-diagonal which implies that the mass operator $\kappa$ in an affine
invariant equation must vanish.

This can be proven for a general finite representation of $SL(4,R)$. Let
us consider the action of a vector operator on an arbitrary irreducible
representation $D(g)$ of $SL(4,R)$ labeled by
$[\lambda_1,\lambda_2,\lambda_3]$, $\lambda_i$ being the number of
boxes in the $i$-th raw,  
\begin{eqnarray*}
  [\lambda_1, \lambda_2, \lambda_3] \otimes [1,0,0] =
  &\,[\lambda_1+1, \lambda_2, \lambda_3] \oplus
  [\lambda_1, \lambda_2+1, \lambda_3] \,\oplus \nonumber \\
  &\,[\lambda_1, \lambda_2, \lambda_3+1]
  \oplus [\lambda_1-1, \lambda_2-1, \lambda_3-1] \,.
\end{eqnarray*}
None of the resulting representations agrees with the representation
$D(g)$ nor with the {\em contragradient} representation $D^{\rm
T}(g^{-1})$ given by
\begin{eqnarray*}
[\lambda_1,\lambda_2,\lambda_3]^{\rm
c}=[\lambda_1,\lambda_1-\lambda_3,\lambda_1-\lambda_2] \,.
\end{eqnarray*}
For a general (reducible) representation this implies vanishing matrices
on the block-diagonal of $X_a$ by similar argumentation as that that led
to the structure of $X_a$. Let the representation space be spanned by
$\Phi=(\varphi_1,\varphi_2,...)^{\rm T}$ with $\varphi_i$ irreducible.
Now we consider the Dirac-type equation in the rest frame
$p_\mu=(E,0,0,0)$ restricted to the subspaces spanned by
$\varphi_i$ $(i=1,2,\dots )$, 
\begin{eqnarray*}
E < {\varphi_i}, X^0  {\varphi_j} > \ =\  
< {\varphi_i}, M {\varphi_i} > \ =\  m_i \delta_{ij} ,
\end{eqnarray*}
where we assumed the operator $M$ to be diagonal. So the mass $m_i$
and therewith $M$ must vanish since $< {\varphi_i}, X^0
{\varphi_i} > = 0$. Therefore, {\bf in an affine invariant Dirac-type wave
equation the mass generation can only be dynamical}, i.e. a result of
an interaction. This agrees with the fact that the Casimir operator
of the special affine group $\overline{SA}(4,R)$ vanishes leaving the
masses unconstrained; thus we expect that our statement also holds for
infinite representations of $\overline{SL}(4,R)$ as well.

\section{$D=3$ Vector operator}

Let us consider now the $\overline{SL}(3,R)$ spinorial representations,
that are necessarily infinite-dimensional. There is a unique
multiplicity-free ("ladder") unitary irreducible representation of the
$\overline{SL}(3,R)$ group,
$D^{(ladd)}_{\overline{SL}(3,R)}(\frac{1}{2})$, that in the reduction
w.r.t. its maximal compact subgroup $\overline{SO}(3)$ yields,
$$
D^{(ladd)}_{\overline{SL}(3,R)}(\frac{1}{2}) \supset
D^{(\frac{1}{2})}_{\overline{SO}(3)} \oplus
D^{(\frac{5}{2})}_{\overline{SO}(3)} \oplus
D^{(\frac{9}{2})}_{\overline{SO}(3)} \oplus \dots
$$
i.e. it has the following $J$ content: $\{J\}$ $=$ $\{\frac{1}{2},\
\frac{5}{2},\ \frac{9}{2},\ \dots\}$.

Owing to the fact that the $\overline{SO}(3)$ and/or $\overline{SL}(3,R)$
vector operators can have nontrivial matrix elements only between the
$\overline{SO}(3)$ states such that $\Delta J = 0, \pm 1$, it is obvious
(on account of the Wigner-Eckart theorem) that all $X$-operator matrix
elements {\it vanish} for the Hilbert space of the
$D^{(ladd)}_{\overline{SL}(3,R)}(\frac{1}{2})$ representation (where
$\Delta J = \pm 2)$. The same holds for the two classes of tensorial
ladder unitary irreducible representations
$D^{(ladd)}_{\overline{SL}(3,R)}(0;\sigma_2)$ and 
$D^{(ladd)}_{\overline{SL}(3,R)}(1;\sigma_2)$, $\sigma_2 \in R$, with the
$J$ content $\{ J \} = \{ 0,2,4,\dots \}$ and 
$\{ J \} = \{ 1, 3, 5,\dots \}$.

Let us consider now the case of $\overline{SL}(3,R)$ unitary irreducible
representations with nontrivial multiplicity w.r.t. the maximal compact
subgroup $\overline{SO}(3)$. An efficient way to construct these
representations explicitly is to set up a Hilbert space of
square-integrable functions $H = L^2([\overline{SO}(3) \otimes
\overline{SO}(3)]^d, \kappa )$, over the diagonal subgroup of the two
copies of the $\overline{SO}(3)$ subgroup, with the group action to the
right defining the group/representation itself while the group action to
the left accounts for the  multiplicity. Here, $\kappa$ stands for the
scalar product kernel, that has to be more singular than the Dirac delta
function in order to account for all types of $\overline{SL}(3,R)$
unitary irreducible representations. We consider the canonical
(spherical) basis of this space $\sqrt{2J+1} D^{J}_{K M} (\alpha , \beta
, \gamma )$, where $J$ and $M$ are the representation labels defined by
the subgroup chain $\overline{SO}(3) \supset \overline{SO}(2)$, while $K$
is the label of the extra copy $\overline{SO}(2)_L \subset
\overline{SO}(3)_L$ that describes nontrivial multiplicity. Here, $-J \le
K,M \le +J$, and for each allowed $K$ one has $J \ge K$, i.e. $J = K,
K+1, K+2, \dots$.

A generic $3$-vector operator ($J=1$) is given now in the spherical basis
($\alpha = 0, \pm 1$) by
$$
X_{\alpha} = {\cal X}_{(0)} D^{(1)}_{0 \alpha} (k) + {\cal X}_{(\pm 1)}
[D^{(1)}_{+1\alpha}(k) + D^{(1)}_{-1\alpha}(k)], \quad k \in
\overline{SO}(3) .
$$
The corresponding matrix elements between the states of two unitary
irreducible $\overline{SL}(3,R)$ representations that are characterized
by the labels $\sigma$ and $\delta$ are given as follows:
\begin{eqnarray*}
&&\left< \begin{array}{c}(\sigma '\ \delta ') \\ J'
\\ K'\ M'\end{array} \right| X_{\alpha} \left|
\begin{array}{c}(\sigma\ \delta ) \\ J \\ K\ M \end{array}
\right> = (-)^{J'-K'} (-)^{J'-M'} \sqrt{(2J'+1)(2J+1)} \\ &&\times\left(
\begin{array}{ccc}\ J' & 1 & J \\ -M' & \alpha & M \\ \end{array}
\right) \Bigg\{
 {{\cal X}_{(0)}}^{(\sigma ' \delta ' \sigma \delta )}_{J'J}
\left(\begin{array}{ccc}\ J' & 1 & J \\ -K' & 0 & K
\end{array} \right) \\
&&+ {{\cal X}_{(\pm 1)}}^{(\sigma ' \delta ' \sigma \delta )}_{J'J}
\left[ \left( \begin{array}{ccc}\ J' & 1 & J \\ -K' & 1 & K \end{array}
\right) + \left( \begin{array}{ccc}\ J' & \ 1 & J \\ -K' & -1 & K
\end{array} \right) \right]\Bigg\}.
\end{eqnarray*}
Therefore, the action of a generic $\overline{SL}(3,R)$ vector operator
on the Hilbert space of some nontrivial-multiplicity unitary irreducible
representation produces the $\Delta J = 0, \pm 1$, as well as the $\Delta
K = 0, \pm 1$ transitions. Owing to the fact that the states of a unitary
irreducible $\overline{SL}(3,R)$ representation are characterized by the
$\Delta K = 0, \pm 2$ condition, it is clear that the $\Delta K = \pm 1$
transitions due to $3$-vector $X$, can take place only between the
states of mutually inequivalent $\overline{SL}(3,R)$ representations whose
multiplicity is characterized by the $K$ values of opposite evenness. In
analogy to the finite-dimensional (tensorial) representation case, the
repeated applications of a vector operator on a given unitary irreducible
(spinorial and/or tensorial) $\overline{SL}(3,R)$ representation would
yield, a priori, an infinite set of irreducible representations. Due to
an increased mathematical complexity of the infinite-dimensional
representations, some additional algebraic constraints imposed on the
vector operator $X$ would be even more desirable than in the
finite-dimensional case. The most natural option is to embed the
$\overline{SL}(3,R)$ $3$-vector $X$ together with the
$\overline{SL}(3,R)$ algebra itself into the (simple) Lie algebra of the
$\overline{SL}(4,R)$ group. Any spinorial (and/or tensorial)
$\overline{SL}(4,R)$ unitary irreducible representation provides now a
Hilbert space that can be decomposed w.r.t. $\overline{SL}(3,R)$ subgroup
representations, and most importantly this space is, by construction,
invariant under the action of the vector operator $X$. Moreover, an
explicit construction of the starting $\overline{SL}(4,R)$
representation generators would yield an explicit form of the $X$ operator.

\section{Embedding into $\overline{SL}(4,R)$}

The $\overline{SL}(4,R)$ group is a $15$-parameter non-compact Lie
group whose defining (spinorial) representation is
given in terms of infinite matrices. All spinorial (unitary and
nonunitary) representations of $\overline{SL}(4,R)$ are necessarily
infinite-dimensional; the finite-dimensional tensorial representations
are nonunitary, while the unitary tensorial representations are
infinite-dimensional. The $\overline{SL}(4,R)$ commutation relations
in the Minkowski space are given by,
$$ [Q_{ab}, Q_{cd}] = i\eta_{bc} Q_{ad} - i\eta_{ad} Q_{cb}.
$$
where, $a,b,c,d = 0,1,2,3$, and $\eta_{ab} = diag(+1, -1, -1, -1)$, while
in the Euclidean space they read,
$$ [Q_{ab}, Q_{cd}] = i\delta_{bc} Q_{ad} - i\delta_{ad} Q_{cb}.
$$
where, $a,b,c,d = 1,2,3,4$, and $\delta_{ab} = diag(+1, +1, +1, +1)$

The relevant subgroup chain reads:
$$
\begin{array}{ccc}
\overline{SL}(4,R) & \supset & \overline{SL}(3,R) \\ \cup &  & \cup \\
\overline{SO}(4), \overline{SO}(1,3) & \supset & \overline{SO}(3),
\overline{SO}(1,2).
\end{array}
$$
We denote by $R_{mn},\ (m,n=1,2,3,4)$ the $6$ compact generators of the
maximal compact subgroup $\overline{SO}(4)$ of the $\overline{SL}(4,R)$
group, and the remaining $9$ noncompact generators (of the
$\overline{SL}(4,R)/\overline{SO}(4)$ coset) by $Z_{mn}$.

In the $\overline{SO}(4) \simeq SU(2)\otimes SU(2)$ spherical basis, the
compact operators are $J^{(1)}_i = \frac{1}{2}(\epsilon_{ijk} R_{jk} +
R_{i4})$ and $J^{(2)}_i = \frac{1}{2}(\epsilon_{ijk} R_{jk} - R_{i4})$,
while the noncompact generators we denote by $Z_{\alpha\beta},\ (\alpha
,\beta = 0, \pm 1)$, and they transform as a $(1, 1)$-tensor operator
w.r.t. $SU(2)\otimes SU(2)$ group. The minimal set of commutation
relations in the spherical basis reads:
\begin{eqnarray*}
&&[J^{(p)}_0, J^{(q)}_{\pm}] = \pm\delta_{pq} J^{(p)}_{\pm}, \quad
[J^{(p)}_+, J^{(q)}_-] = 2\delta_{pq} J^{(p)}_0, \quad (p,q=1,2), \\
&&[J^{(1)}_0, Z_{\alpha\beta}] = \alpha Z_{\alpha\beta}, \quad
[J^{(1)}_{\pm}, Z_{\alpha\beta}] = \sqrt{2 - \alpha (\alpha \pm 1)}
Z_{\alpha \pm 1\ \beta} \\ &&[J^{(2)}_0, Z_{\alpha\beta}] = \beta
Z_{\alpha\beta}, \quad [J^{(2)}_{\pm}, Z_{\alpha\beta}] = \sqrt{2 - \beta
(\beta \pm 1)} Z_{\alpha \beta\pm 1} \\ && [Z_{+1\ +1}, Z_{-1\ -1}] = -
(J^{(1)} + J^{(2)}).
\end{eqnarray*}
The $\overline{SO}(3)$ generators are $J_i = \epsilon_{ijk}J_{jk},\
J_{ij}\equiv R_{ij},\  (i,j,k =1,2,3)$, while the traceless
$T_{ij}=Z_{ij}\ (i,j=1,2,3)$ define the coset
$\overline{SL}(3,R)/\overline{SO}(3)$. In the $\overline{SO}(3)$
spherical basis the compact operators are $J_0,\ J_{\pm 1}$, while the
noncompact ones $T_{\rho}, \ (\rho = 0, \pm 1, \pm 2)$ transform w.r.t.
$\overline{SO}(3)$ as a quadrupole operator. The corresponding minimal
set of commutation relations reads:
\begin{eqnarray*}
&&[J_0, J_{\pm}] = \pm J_{\pm}, \quad [J_+, J_- ] = 2J_0 \\ && [T_{+2},
T_{-2}] = - 4J_0.
\end{eqnarray*}

There are three (independent) $\overline{SO}(3)$ vectors in the algebra
of the $\overline{SL}(4,R)$ group. They are: the $\overline{SO}(3)$
generators themselves, $N_i \equiv R_{i4} = Q_{i0} + Q_{0i}$, and $K_i
\equiv Z_{i4} = Q_{i0} - Q_{0i}$. From the latter two, one can form the
following linear combinations,
$$
A_i = \frac{1}{2}(N_i + K_i) = Q_{i0}, \quad B_i = \frac{1}{2}(N_i - K_i)
= Q_{0i}.
$$
The commutation relations between $N$, $K$, $A$, and $B$ and the
$\overline{SL}(3,R)$ generators read:
\begin{eqnarray*}
&&{}[J_i, N_j] = i\epsilon_{ijk}N_k, \quad [T_{ij}, N_k] =
i(\delta_{ik}K_j + \delta_{jk}K_i), \\ &&{}[J_i, K_j] =
i\epsilon_{ijk}K_k, \quad [T_{ij}, K_k] = i(\delta_{ik}N_j +
\delta_{jk}N_i), \\ &&{}[J_i, A_j] = i\epsilon_{ijk}A_k, \quad [T_{ij},
A_k] = i(\delta_{ik}A_j + \delta_{jk}A_i), \\ &&{}[J_i, B_j] =
i\epsilon_{ijk}B_k, \quad [T_{ij}, B_k] = -i(\delta_{ik}B_j +
\delta_{jk}B_i).
\end{eqnarray*}
It is clear from these expressions that only $A_i$ and $B_i$ are
$\overline{SL}(3,R)$ vectors as well. More precisely, $A$ transforms
w.r.t. $\overline{SL}(3,R)$ as the $3$-dimensional representation
$[1,0]$, while $B$ transforms as its contragradient $3$-dimensional
representation $[1,1]$.

To summarize, either of the choices
$$
X_i \sim A_i, \quad X_i \sim B_i
$$
insures that a Dirac-like wave equation $(iX\partial - m)\Psi (x) = 0$
for a (infinite-component) spinor field is fully $\overline{SL}(3,R)$
covariant. The choices
$$
X_i \sim J_i, \quad X_i \sim N_i, \quad X_i \sim K_i ,
$$
would yield wave equations that are Lorentz covariant only; though the
complete $\overline{SL}(3,R)$ acts invariantly in the space of $\Psi (x)$
components. It goes without saying that the correct unitarity properties
can be accounted for by making use of the deunitarizing automorphism, as
discussed above.

Due to complexity of the generic unitary irreducible representations
of the $\overline{SL}(4,R)$ group, we confine here to the
multiplicity-free case only. In this case, there are just two, mutually
contragradient, representations that contain spin $J = \frac{1}{2}$
representation of the $\overline{SO(3)}$ subgroup, and belong to the
set of the so called Descrete Series i.e.
$$ 
D^{disc}_{\overline{SL}(4,R)}(\frac{1}{2},0) \supset
D^{(\frac{1}{2},0)}_{\overline{SO}(4)} \supset 
D^{\frac{1}{2}}_{\overline{SO}(3)} \quad 
D^{disc}_{\overline{SL}(4,R)}(0,\frac{1}{2}) \supset
D^{(0,\frac{1}{2})}_{\overline{SO}(4)} \supset 
D^{\frac{1}{2}}_{\overline{SO}(3)} .
$$
The full reduction of these representations to the representations of
the $\overline{SL}(3,R)$ subgroup reads:
\begin{eqnarray*}
D^{disc}_{\overline{SL}(4,R)}(j_0,0) &\rightarrow&
{\Sigma^{\oplus}}_{j=1}^{\infty}
D^{disc}_{\overline{SL}(3,R)}(j_0;\sigma_2,\delta_1,j) \\
D^{disc}_{\overline{SL}(4,R)}(0,j_0) &\rightarrow&
{\Sigma^{\oplus}}_{j=1}^{\infty}
D^{disc}_{\overline{SL}(3,R)}(j_0;\sigma_2,\delta_1,j)
\end{eqnarray*}

Finally, by making use of the known expressions of the
$\overline{SL}(4,R)$ generators matrix elements for these spinorial
representations, we can write down an $\overline{SL}(3,R)$ covariant
wave equation in the form
\begin{eqnarray*}
&&(iX^{\mu}\partial_{\mu} - M) \Psi (x) = 0 , \\ && \Psi\ \sim\
D^{disc}_{\overline{SL}(4,R)}(\frac{1}{2},0) \oplus
D^{disc}_{\overline{SL}(4,R)}(0,\frac{1}{2}) ,\\
&&X^{\mu} \in \{ Q^{\mu 0},\  Q^{0 \mu} \} .
\end{eqnarray*}

\vfill\break

\end{document}